\documentclass{jetpl}
\twocolumn

\lat

\title{Signatures of Quantum Chaos in Wave Functions Structure for Multi-well 2D Potentials}

\rtitle{Signatures of Quantum Chaos in Wave Functions
Structure\ldots}

\sodtitle{Signatures of Quantum Chaos in Wave Functions Structure
for Multi-well 2D Potentials}

\author{V.\,P.\,Berezovoj$^+$,
Yu.\,L.\,Bolotin$^+$, V.\,A.\,Cherkaskiy$^+$\/\thanks{e-mail:
cherkaskiy@kipt.kharkov.ua}}

\rauthor{V.\,P.\,Berezovoj, Yu.\,L.\,Bolotin, V.\,A.\,Cherkaskiy}

\sodauthor{Berezovoj, Bolotin, Cherkaskiy}

\address{$^+$ A.I.Akhiezer Institute for Theoretical Physics, National Science
Center "Kharkov Institute of Physics and
Technology", Akademicheskaya Str. 1, 61108 Kharkov, Ukraine\\~\\
}

\dates{1 November 2003}{*}

\abstract{We propose a new approach to investigation of quantum
manifestations of classical stochasticity (QMCS) in wave functions
structure, which can be realized in potentials with two and more
local minima. The main advantage of the proposed approach is the
possibility to detect QMCS in comparison not different wave
functions, but different parts of the same wave function.
Efficiency of the approach is demonstrated for two potentials:
surface quadrupole oscillations (QO) and lower umbillic
catastrophe (UC) $D_5$.}

\PACS{05.45.Mt, 05.45.Pq}

\begin{document}

\maketitle

\begin{figure}
%%%%%%%%%%%%%%%%%%%%%%%%%%%%%%%%%%%%%%%%%%%%%%%%%%%%%%%%%% fig 2 %%%%%%%%%%%%%%%%%
\includegraphics[width=0.5\textwidth]{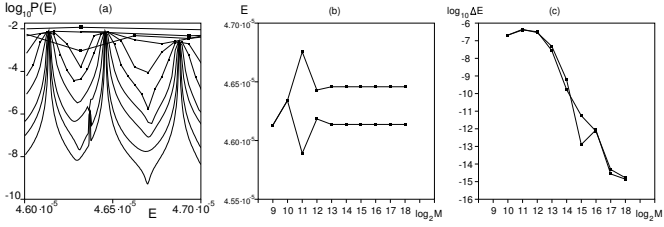}
\caption{Figure 2: Determination of two close energy levels $E_n$
and $E_{n+1}$ in the QO potential (\ref{Uxydl}) with $W=18$ and
different number of time steps $M=2^k, k = 9,10...18$: (a) -
absolute value of $P^{(k)}(E)$ in log scale, (b) - values of
$E^{(k)}_n$ and $E^{(k)}_{n+1}$, (c) - consecutive corrections
$E^{(k-1)}_n-E^{(k)}_n$ and
$E^{(k-1)}_{n+1}-E^{(k)}_{n+1}$.\label{satur}}
\end{figure}

Energy spectra and eigenfunctions of classically non-integrable
systems represent the main object of search for QMCS
\cite{gutzwiller,haake,stockmann}. It should be pointed out that
in analysis of QMCS in the energy spectra the principal role was
given to statistical characteristics, i.e. quantum chaos was
treated as property of a group of states. In contrast, the choice
of a stationary wave function as a basic object of investigation,
relates quantum chaos to an individual state. Usual procedure of
search for QMCS in wave function implies investigation of
distinction in its structure below and above the classical energy
of transition to chaos (or other parameters of regularity-chaos
transition). Such procedure meets difficulties connected with
necessity to separate QMCS from modifications of wave functions
structure due to trivial changes in its quantum numbers. Up to
present time correlations between peculiarities of the classical
motion and structure of wave functions were studied mostly for
billiard-type systems \cite{berry77, mcdonald, robnik}. For
Hamiltonian systems with non-zero potential energy QMCS were
studied either for model wave functions \cite{heller} or for
potential energy surfaces (PES) with simple geometry
\cite{barbanis}. Till now there is practically no information on
wave functions structure for generic Hamiltonian systems,
including multi-well potentials. Such systems allow existence of
the mixed state (MS): different (regular or chaotic) classical
regimes coexist in different local minima at fixed energy
\cite{yadfiz,obzor}. Aim of our work is to show, that such systems
represent optimal object for investigation of QMCS in wave
functions structure. Wave functions of MS allow to find QMCS in
comparison not different eigenfunctions, but different parts of
the same wave function, situated in different regions of
configuration space (corresponding to different local minima of
the potential). \DeclareGraphicsRule{.gif}{eps}{.gif.bb}{'gif2eps
#1'}
\begin{figure}
%%%%%%%%%%%%%%%%%%%%%%%%%%%%%%%%%%%%%%%%%%%%%%%%%%%%%%%%%% fig 1 %%%%%%%%%%%%%%%%%
\includegraphics[width=0.5\textwidth]{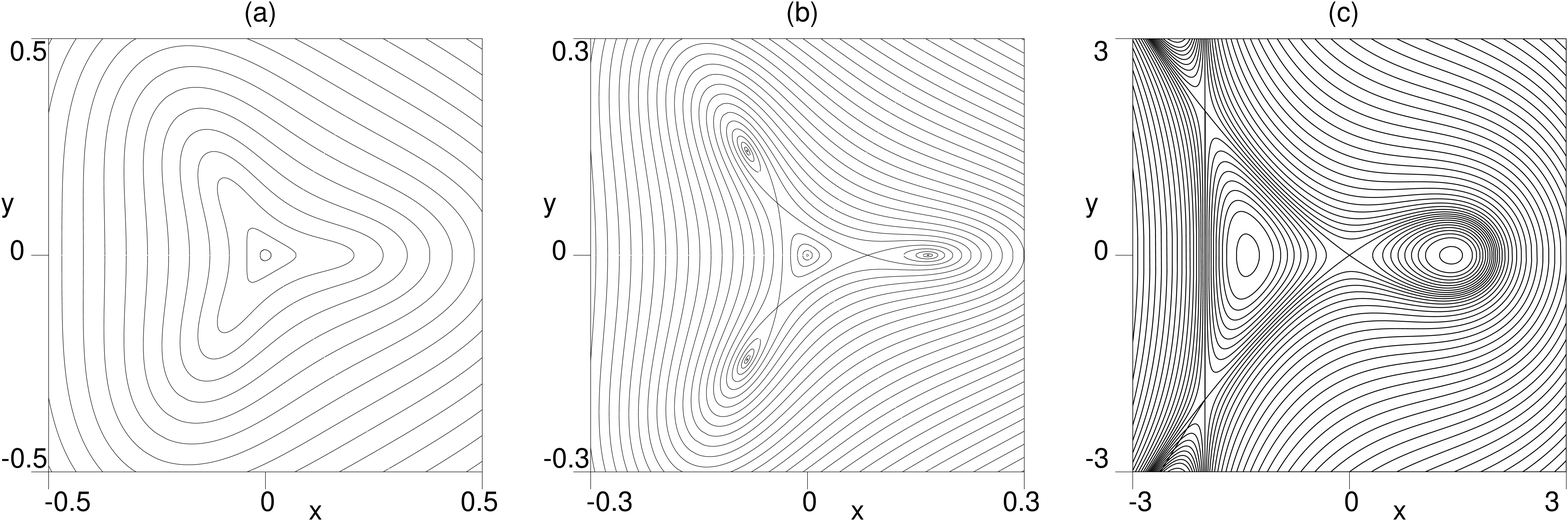}
\caption{Figure 1: The level lines of the QO potential
(\ref{Uxydl}) for $W=13$ (a), $W=18$ (b), and for the UC $D_5$
(\ref{d5}) with $a = 2$ (c). \label{w13w18d5}}
\end{figure}
\\ Let us demonstrate this possibility for MS,
generated by the deformation potential of surface QO of atomic
nuclei \cite{eizenberg} and lower UC $D_5$ \cite{catastrophe}. It
can be shown \cite{mozel}, that using only the transformation
properties of the interaction, the  QO potential takes the form
\begin{equation}
\label{Uaa} U_{QO}(a_0,a_2)=\sum_{m,n}C_{mn}(a_0^2+2a_2^2)^m a_0^n
(6a_2^2-a_0^2)^n
\end{equation}
where $a_0$ and $a_2$ are internal coordinates of nuclear surface
undergoing the QO
\begin{equation}
\label{Rtf} R(\theta,\varphi)=R_0\{1+a_0
Y_{2,0}(\theta,\varphi)+a_2
[Y_{2,2}(\theta,\varphi)+Y_{2,-2}(\theta,\varphi)]\}
\end{equation}
Since in the construction of (\ref{Uaa}) only transformation
properties of interaction play role, this expression describes
potential energy of surface QO of a charged liquid drop of any
nature (for example, a metal cluster \cite{saunders}), containing
specific character of the interaction only in the coefficients
$C_{mn}$. Restricting ourselves with the terms of fourth order in
the deformation and assuming equality of masses for the two
independent directions, we get the following $C_{3v}$-symmetric
Hamiltonian
\begin{equation}
\label{H=T+U} H = \frac{p_x^2+p_y^2}{2m}+U_{QO}(x,y;a,b,c)
\end{equation}
where
\begin{equation}
\label{Uxy}
\begin{array}{c}
U_{QO}(x,y;a,b,c) =\\ \frac{a}{2}(x^2+y^2)+b\left(xy^2-\frac 1 3
x^3\right) +c\left(x^2+y^2\right)^2\\
x = a_0, y = \sqrt{2}a_2, a = 2C_{10}, b = 3C_{01}, c = C_{20}
\end{array}
\end{equation}
\\Let us introduce the dimensionless variables
\begin{subequations}
\begin{equation}
\label{xy-dimless} (x,y)=l_0(\bar{x},\bar{y}),
(p_x,p_y)=p_0(\bar{p}_x,\bar{p}_y), E = \varepsilon_0\bar{E}
\end{equation}
\begin{equation}
\label{scales} l_0=\frac{b}{c}, p_0=\sqrt{m\frac{b^4}{c^3}},
\varepsilon_0 = \frac{b^4}{c^3}
\end{equation}
\end{subequations}
In the variables $(\bar{x},\bar{y})$ (further we will drop the bar
line) the Hamiltonian (\ref{H=T+U}) has the form
\begin{subequations}
\begin{equation}
\label{Hxydl} H = \frac{p_x^2+p_y^2}{2m}+U_{QO}(x,y;W)
\end{equation}
\begin{equation}
\label{Uxydl} U_{QO}(x,y;W) = \frac{1}{2W}(x^2+y^2)+xy^2-\frac 1 3
x^3+\left(x^2+y^2\right)^2
\end{equation}
\end{subequations}
Hamiltonian (\ref{Hxydl}) and corresponding equations of motion
depend only on $W=b^2/(ac)$, which is the unique dimensionless
parameter, that can be constructed from $a,b$ and $c$, and it
completely determines the PES [Fig.\ref{w13w18d5}(a),(b)]. Region
$0<W\le16$ includes potentials with only one critical point
--- minimum in the origin [Fig.\ref{w13w18d5}(a)], corresponding to
spherically symmetric equilibrium shape of the nucleus (or liquid
charged drop). For $W>16$ the PES has seven critical points: four
minima (one central and three peripheral) and three saddles,
separating the peripheral minima from the central one
[Fig.\ref{w13w18d5}(b)]. In this Letter we consider in details the
case $W=18$, when the potential (\ref{Uxydl}) has four minima with
the same value $E_{min}=0$ and the saddle energies $E_S=1/20736$.
It was shown \cite{yadfiz}, that the critical energy of transition
to chaos $E_{cr}$ has different values for different minima:
$E_{cr}=E_S/2$ for the central minimum and $E_{cr}=E_S$ for the
peripheral ones. It means, that for $E_S/2<E<E_S$ regular and
chaotic trajectories coexist and are separated not in phase, but
in configuration space, resulting in the phenomenon of MS.
\\ MS is a common case for multi-well potentials.
According to catastrophe theory, a wide class of multi-well 2D
polynomial potentials can be generated by germs of lower UC of
types $D_5,D^\pm_6,D_7$ from the Thom catastrophes list, affected
by certain perturbation \cite{catastrophe}. To demonstrate the
proposed approach we consider, apart from the QO potential
(\ref{Uxydl}), a lower UC $D_5$, described by germ $x^4/4+y^2x$
with perturbation $bx^2-ay^2$. This potential has only two local
minima and three saddles [Fig.\ref{w13w18d5}(c)], and therefore it
is the simplest potential, where MS is observed. Under the Maxwell
condition $b=a^2/4$ the energies of all the saddles are the same,
and energies of all the local minima too. We will consider the
case $a=2,b=1$
\begin{equation} \label{d5} U_{D_5}(x,y) =
\frac{x^4}{4}+xy^2+2y^2-x^2
\end{equation}
when all $E_{min}=-1$ and all $E_S=0$, and MS is observed in the
energy region $-1/2<E_{MS}<0$.
\\
Calculation of quasiclassical part of the spectrum for systems
with multi-well PES requires appropriate numerical methods. Matrix
diagonalization (MD) method is attractive for the Hamiltonians
with eigenfunctions, that do not differ too much from the basis
 functions. However this numerical procedure
becomes less attractive (or even not efficient at all) at the
transition to PES of complicated topology (multi-well potentials).
In particular, the diagonalization of the QO Hamiltonian
(\ref{Hxydl}) with $W>16$ in the harmonic oscillator basis
requires so large number of the basis functions, that goes beyond
the limits of the computation power. In this case the attractive
alternative to the MD may become the spectral method (SM) in the
form, proposed by Feit et al. \cite{spectral method}. \\Numerical
solution of the stationary Schr\"odinger equation
\begin{equation}
\label{se}
\left[-\frac{\hbar^2}{2}\Delta+U(x,y)\right]\psi_n(x,y)=E_n\psi_n(x,y)
\end{equation}
by the SM requires computation of the correlation function
\begin{equation}
\label{ptbd} P(t)=\int dxdy\psi_0^*(x,y)\psi(x,y,t)
\end{equation}
where $\psi(x,y,t)$ represents a numerical solution of the
corresponding time-dependent Schr\"odinger equation with an
arbitrary initial condition $\psi_0(x,y)=\psi(x,y,t=0)$. The
solution $\psi(x,y,t)$ can be accurately generated with the help
of the split operator method
\begin{equation}
\begin{array}{c}
\psi(x,y,t+\Delta t)=\\e^{i\frac{\hbar^2\Delta
t\nabla^2}{4}}e^{-i\Delta t U(x,y)}e^{i\frac{\hbar^2\Delta
t\nabla^2}{4}}\psi(x,y,t)+O(\Delta t^3)
\end{array}
\end{equation}
where $\exp(i{\hbar^2\Delta t\nabla^2}/{4})\psi(x,y,t)$ is
evaluated with the help of the band-limited Fourier series
representation
\begin{equation}
\psi(x,y,t)=\sum_{m=-N/2+1}^{N/2}\sum_{n=-N/2+1}^{N/2}\psi_{mn}(t)e^{\frac{2\pi
i}{L_0}(mx+ny)}
\end{equation}
where $N$ is the number of grid points along a grid line and $L_0$
is the grid length. \\From the other hand, the solution
$\psi(x,y,t)$ can be expressed as a linear superposition of
eigenfunctions $\psi_n(x,y)$ of (\ref{se})
\begin{equation}
\label{psit} \psi(x,y,t)=\sum_na_n\psi_n(x,y)\exp(-iE_nt/\hbar)
\end{equation}
We assume no degenerate states in the decomposition (\ref{psit}),
which can always be achieved by certain choice of the initial
condition $\psi_0(x,y)$. Using (\ref{psit}) in (\ref{ptbd}), we
obtain
\begin{equation}
\label{pt} P(t)=\sum_n|a_n|^2\exp(-iE_nt / \hbar)
\end{equation}
The Fourier transform of (\ref{pt})
\begin{equation}
\label{pe}
 P(E)=\frac 1 T
\int\limits_0^Tdte^{i\frac{Et}{\hbar}}P(t)w(t)=\sum_n|a_n|^2\delta_T(E-E_n)
\end{equation}
where
\begin{equation}
\begin{array}{c}
\delta_T(E)=\frac 1 T\int\limits_0^Tdtw(t)e^{i\frac{Et}{\hbar}}=\\
\frac{e^{i\frac{ET}{\hbar}}-1}{i\frac{ET}{\hbar}}- \frac 1 2
\left[
\frac{e^{i(\frac{ET}{\hbar}+2\pi)}-1}{i(\frac{ET}{\hbar}+2\pi)}+
\frac{e^{i(\frac{ET}{\hbar}-2\pi)}-1}{i(\frac{ET}{\hbar}-2\pi)}
 \right]
 \end{array}
\end{equation}
and $w(t)=1-\cos(2\pi t/T)$ is the Hanning window function. The
plot for $P(E)$ displays a set of sharp local maxima at $E=E_n$
[Fig.\ref{satur}(a)], where $E_n$ are the energy eigenvalues of
(\ref{se}).
Once the eigenvalues are known, the corresponding eigenfunctions
can be computed by numerically evaluating the integrals
\begin{equation}
\psi_n(x,y)=\frac 1
T\int\limits_0^Tdt\psi(x,y,t)w(t)e^{i\frac{E_nt}{\hbar}}
\end{equation}
This procedure is very efficient, when implemented with the help
of the fast Fourier transform algorithm (FFT), and very accurate,
since the spatial derivatives are approximated to $N$th order in
$\Delta x=\Delta y=L_0/N$. $L_0$ and $N$ must be chosen large
enough, and $\psi_0(x,y)$ sufficiently fast decaying with its
Fourier components, in order to assure that $\psi(x,y,t)$ is
negligible on the grid boundaries both in coordinate and
reciprocal spaces. The sampling interval $\Delta t$ limits the
spectral bandwidth $\Delta E_{max}=\pi\hbar/\Delta t$ of a
function that can be represented by a Fourier series determined by
sampled values. Therefore $\Delta t$ should be chosen small enough
to accommodate the necessary number of energy levels, or $\Delta
E_{max}>\Delta U_{max}$, where $\Delta U_{max}$ is the maximum
excursion of the potential. Since the potential (\ref{Uxydl}) is
unbounded, it is necessary to put an appropriate cutoff in order
to apply the SM. If the wave function $\psi(x,y,t)$ is generated
over a total time $T=M\Delta t$, the minimum separation in energy
levels that can be resolved is $\Delta E_{min} = \pi\hbar/T$,
which also provides an estimate of the accuracy with which
individual eigenvalues can be determined from the numerically
computed $P(E)$ (\ref{pe}) without the aid of lineshape fitting
techniques. Such techniques, however, may improve the eigenvalues
accuracy by roughly two orders of magnitude. Fig.\ref{satur} shows
the typical shape of $P(E)$ [Fig.\ref{satur}(a)] and the
corresponding two close energy levels [Fig.\ref{satur}(b)],
calculated for increasing $M=2^k, k=9, 10, ... 18$, using the
single-line fit. For $k=9,10$ the levels look as one, at $k\ge 11$
they are resolved, and for $k>13$ their values change much less
than the level spacing. Fig.\ref{satur}(c) shows the differences
between the eigenvalues, calculated with given $M=2^k$ and the
more accurate, calculated with $M=2^{k+1}$ --- the procedure
converges fast indeed, displaying the saturation in accuracy.
\\In applying the SM, a preponderant fraction of the computer time
goes to the generation of the $\psi(x,y,t)$, and most of that time
in turn is invested in FFT computations, so the overall
calculation time for our 2D problems scales as $MN^2\ln{N}$. We
performed our calculations with $N=512$ and $M=65 536$, which
allowed to obtain for reasonable time about $10^2$ eigenfunctions
with sufficient accuracy and high coordinate resolution to allow
their detailed analysis. The scaled Planck's constant $\hbar$ is
an arbitrary parameter and is chosen to obtain the desired number
of energy levels.
\begin{figure*}
%%%%%%%%%%%%%%%%%%%%%%%%%%%%%%%%%%%%%%%%%%%%%%%%%%%%%%%%%% fig 3 %%%%%%%%%%%%%%%%%
\parbox{0.5\textwidth}{
\includegraphics[width=0.5\textwidth]{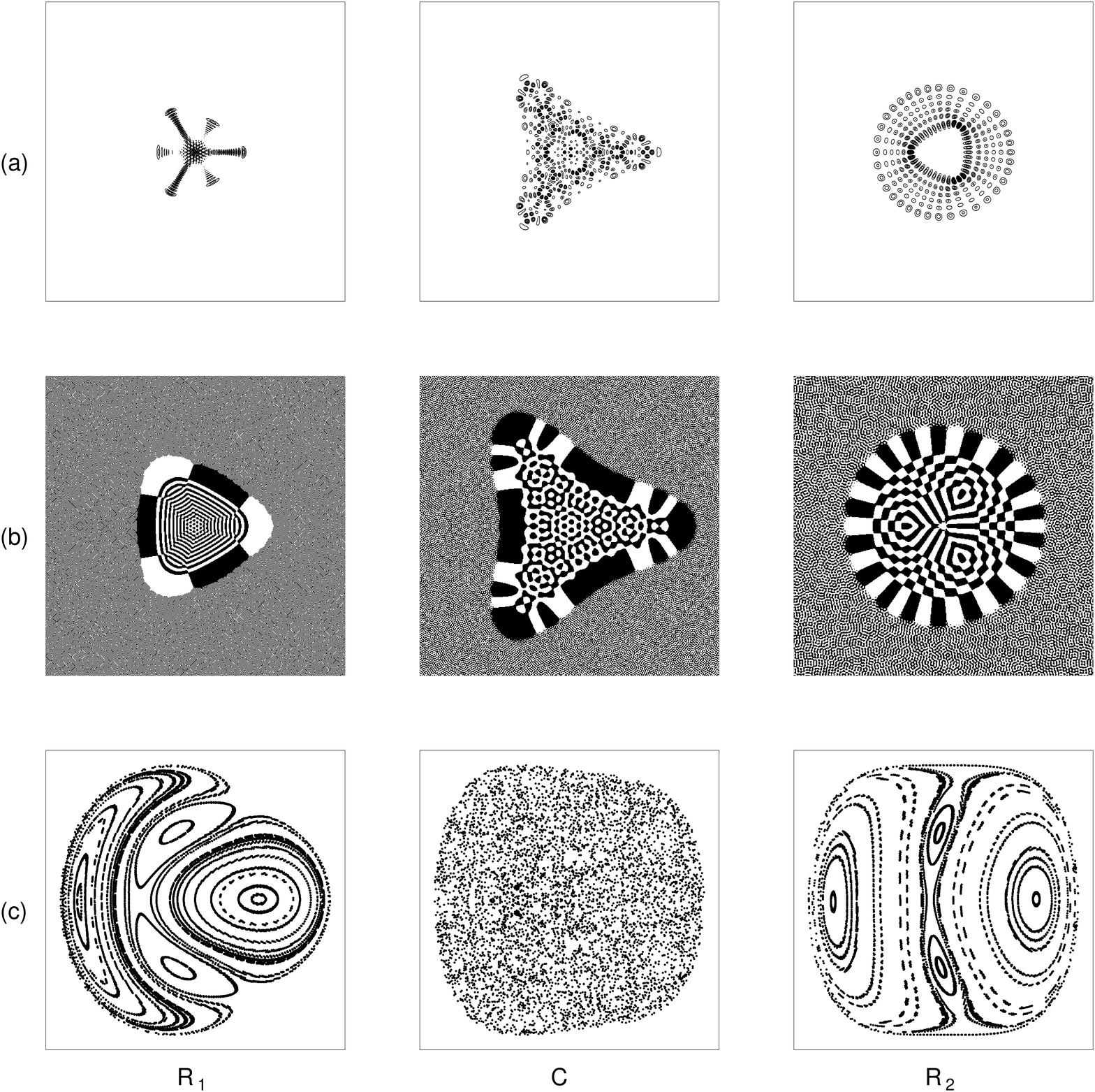}
%%%%%%%%%%%%%%%%%%%%%%%%%%%%%%%%%%%%%%%%%%%%%%%%%%%%%%%%%% fig 4 %%%%%%%%%%%%%%%%%
\caption{Figure 3: The R-C-R transition in the QO potential
(\ref{Uxydl}) with $W=13$. \label{w13}} }
\parbox{0.5\textwidth}{
\includegraphics[width=0.5\textwidth]{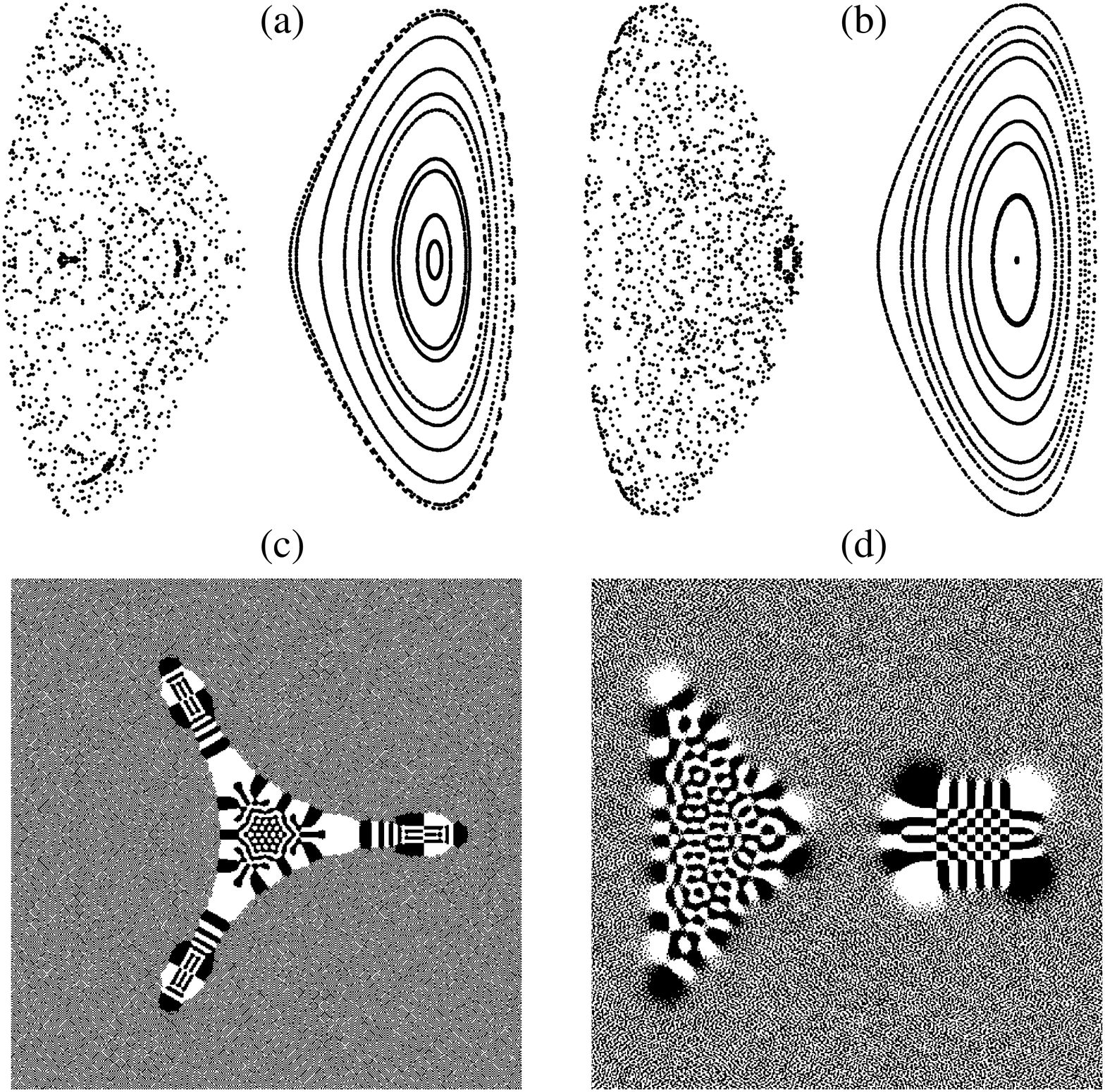}
\caption{Figure 4: The MS in the QO potential (a),(c) and in the
$D_5$ UC (b),(d): (a),(b) -- Poincar\'e surfaces of section,
(c),(d) -- nodal domains of the eigenfunctions. \label{w18d5}} }
\end{figure*}
\\Let us now discuss the results obtained for the potentials
$U_{QO}$ and $U_{D_5}$. In the former case for $W<16$ the only
possibility to detect the QMCS in the wave function structure is
to look how it changes as energy grows. As was shown in
\cite{old}, for $4<W<16$ energy region of chaotic motion is
bounded from both sides: $E_{cr1}<E_{c}<E_{cr2}$, which means the
regularity-chaos-regularity (R-C-R) transition at energies
$E_{cr1}$ and $E_{cr2}$ respectively. We will distinguish three
energy regions: low-energy regular $R_1 (E<E_{cr1})$, chaotic $C
(E_{cr1}<E<E_{cr2})$ and high-energy regular $R_2 (E>E_{cr2})$.
For the considered case $W=13$ the R-C-R transition is observed at
critical energies $E_{cr1}=8\cdot10^{-5}$ and $E_{cr2}=8.4\cdot
10^{-2}$ respectively. Fig.\ref{w13} shows the changes in the
structure of level lines of $|\psi_n|^2$ [Fig.\ref{w13}(a)] and
the corresponding nodal domains picture [Fig.\ref{w13}(b)],
clearly correlating with the character of the classical motion,
displayed in the corresponding Poincar\'e surfaces of section
[Fig.\ref{w13}(c)].
\\The considered possibility corresponds to the traditional
approach in search of QMCS in the wave function structure.
Existence of the MS, at $W>16$ for the QO potential
[Fig.\ref{w18d5}(a)] or in the UC $D_5$ potential
[Fig.\ref{w18d5}(b)], opens a new possibility. Comparing the
structure of the eigenfunction in central and peripheral minima of
the QO potential [Fig.\ref{w18d5}(c)], or in left and right minima
of the UC $D_5$ potential [Fig.\ref{w18d5}(d)], it is evident that
the nodal structures of the regular part and the chaotic part of
the eigenfunction are clearly different:
\\ i) within the classically allowed region the nodal
domains of the regular part of the wave function form a well
recognizable checkerboard-like pattern \cite{gutzwiller}; nothing
similar can be observed for the chaotic part;
\\ ii) the nodal lines of the regular part exhibit crossings or very
tiny quasicrossings; in the chaotic part the nodal lines
quasicrossings have significantly larger avoidance ranges;
\\ iii) while crossing the classical turning line $U(x,y)=E_n$, the
nodal lines structure of the regular part immediately switches to
the straight nodal lines, going to infinity, which makes the
turning point line itself easily locatable in the nodal domains
structure; in the chaotic part an intermediate region exists
around the turning line, where some of the nodal lines pinch-off,
making transition to the classically forbidden region more
graduate and not so manifesting in the nodal structure.
\\In conclusion we remark that the Hamiltonian system with
multi-well PES represents a realistic model, describing the
dynamics of transitions between different equilibrium states,
including such important cases, as chemical reactions and nuclear
fission. Existence of the MS must essentially determine dynamics
of the physical processes in such systems, for instance the
quantum assistance tunnelling. We demonstrated a possibility to
observe the QMCS in an individual quantum mechanical state --- an
eigenfunction of the MS. Further analysis implies investigation of
the eigenfunction amplitude distribution, nodal lines
quasicrossings avoidance range distribution, and the wave packets
dynamics. Another interesting perspective is to study relevance of
the Berry-Robnik formula for the energy levels spacings
distribution in the MS and to investigate the nodal domains and
the nodal lines statistics.

\end{document}